\documentclass[conference]{IEEEtran}
\IEEEoverridecommandlockouts
\usepackage{cite}
\usepackage{amsmath,amssymb,amsfonts}
\usepackage{algorithmic}
\usepackage{graphicx}
\usepackage{textcomp}
\usepackage{xcolor}
\def\BibTeX{{\rm B\kern-.05em{\sc i\kern-.025em b}\kern-.08em
    T\kern-.1667em\lower.7ex\hbox{E}\kern-.125emX}}
\begin{document}

\title{
Eye-tracking in Mixed Reality  for Diagnosis of Neurodegenerative Diseases  \\
\thanks{The project was funded by The National Centre for Research and Development, Poland under Lider Grant no: LIDER/6/0049/L-12/20/NCBIR/2021.}
}

\author{Mateusz Daniol$^{1}$, Daria Hemmerling$^{1}$, Jakub Sikora$^{1}$, Pawel Jemiolo$^{1}$, \\ Marek Wodzinski$^{1,2}$,  and Magdalena Wojcik-Pedziwiatr$^{3}$
\thanks{The project was funded by The National Centre for Research and Development, Poland under Lider Grant no: 0049/L-12/2020.}
\thanks{$^{1}$Mateusz Daniol, Daria Hemmerling, Jakub Sikora, Pawel Jemiolo, and Marek Wodzinski are with the AGH University of Kraków, Faculty of Electrical Engineering, Automatics, Computer Science, and Biomedical Engineering, Krakow, Poland.
        {\tt\small hemmer@agh.edu.pl}}%
\thanks{$^{2}$Marek Wodzinski is also with the University of Applied Sciences Western Switzerland (HES-SO Valais), Information Systems Institute, Sierre, Switzerland.}%
\thanks{$^{3}$Magdalena Wojcik-Pedziwiatr is with the Andrzej Frycz Modrzewski Krakow University, Department of Neurology, Krakow, Poland.

© 2024 IEEE must be obtained for all other uses, in any current or future media, including reprinting/republishing this material for advertising or promotional purposes, creating new collective works, for resale or redistribution to servers or lists, or reuse of any copyrighted component of this work in other works.}%
}

\maketitle

\begin{abstract}

Parkinson's disease ranks as the second most prevalent neurodegenerative disorder globally. This research aims to develop a system leveraging Mixed Reality capabilities for tracking and assessing eye movements. In this paper, we present a medical scenario and outline the development of an application designed to capture eye-tracking signals through Mixed Reality technology for the evaluation of neurodegenerative diseases. Additionally, we introduce a pipeline for extracting clinically relevant features from eye-gaze analysis, describing the capabilities of the proposed system from a medical perspective. The study involved a cohort of healthy control individuals and patients suffering from Parkinson's disease, showcasing the feasibility and potential of the proposed technology for non-intrusive monitoring of eye movement patterns for the diagnosis of neurodegenerative diseases.
\newline

\indent \textit{Clinical relevance}— Developing a non-invasive biomarker for Parkinson's disease is urgently needed to accurately detect the disease's onset. This would allow for the timely introduction of neuroprotective treatment at the earliest stage and enable the continuous monitoring of intervention outcomes. The ability to detect subtle changes in eye movements allows for early diagnosis, offering a critical window for intervention before more pronounced symptoms emerge. Eye tracking provides objective and quantifiable biomarkers, ensuring reliable assessments of disease progression and cognitive function. The eye gaze analysis using Mixed Reality glasses is wireless, facilitating convenient assessments in both home and hospital settings. The approach offers the advantage of utilizing hardware that requires no additional specialized attachments, enabling examinations through personal eyewear. 

\end{abstract}


\section{Introduction}

Parkinson's disease (PD) is a prevalent neurodegenerative disorder affecting over 10 million individuals worldwide \cite{pd10}. Identifying and characterizing biomarkers for PD have become increasingly important for early diagnosis and effective disease monitoring. Recently, the analysis of eye movements became a significant research field among the potential biomarkers, mostly due to the potential to provide insights into the neuronal mechanisms and pathways involved in PDF pathogenesis.
In Mixed Reality (MR) glasses, modern eye-tracking technology uses infrared light sources and cameras to monitor the wearer's eye movement continuously. This hardware configuration captures pupil and corneal reflections, providing precise data on gaze direction. The real-time processing unit in MR glasses computes the wearer's gaze point, enabling interactions like seamless object selection, menu navigation, and dynamic adjustment of virtual elements based on gaze. Reflective waveguide technology, using mirrors to redirect light, offers an efficient solution for developing highly effective eye-tracking systems.


Ocular motor disruptions in PD involve various abnormalities, particularly affecting reflexive saccades (RS), antisaccades (AS), memory-guided saccades (MGS), and smooth pursuit (SP). In PD, RS are often hypometric, especially vertically, indicating a specific disruption in saccadic eye movement. AS tasks reveal increased latency and error rates, serving as a distinctive marker from other disorders. PD-related impairment in MGS is characterized by reduced accuracy and increased latency, distinguishing it from conditions like Huntington’s disease. Additionally, SP in PD shows reduced velocity and smoothness, with distinctions from cerebellar ataxia or multiple system atrophy.

The exploration of eye movements as a potential biomarker for the diagnosis and measurement of Parkinson's disease has demonstrated encouraging results \cite{sun_monitoring_2023, jung_abnormal_2019, alfalahi_scoping_2023, krichen_anomalies_2021, aarsland_parkinson_2021, motahari-nezhad_digital_2022}. Furthermore, the non-invasive, objective, and rapid nature of eye tracking makes it an attractive approach for evaluating oculomotion, with the added potential for gamification mechanisms using MR technology to enhance diagnostic capabilities. In the area of MR in medicine, Microsoft Hololens (1 \& 2) glasses are the main player, but from over 200 applications reviewed by Gsaxner et al. only a few are related to patient diagnostics \cite{gsaxner_hololens_2023}. What is more important there are almost no studies evaluating the use of MR glasses in monitoring neurodegenerative diseases \cite{orlosky_emulation_2017}. There is also a lack of studies investigating off-the-shelf eye tracking in detecting PD-related eye movement disorders.

Traditional methods of eye movement monitoring face several challenges, including the requirement for costly equipment, specialized software, and difficulties in adapting them for use in a hospital environment \cite{jung_abnormal_2019, kourtis_digital_2019}. 
Our study using the Microsoft HoloLens 2 (HL2) tackles cost and complexity issues by leveraging its integrated eye-tracking features. This user-friendly solution eliminates the need for separate, expensive eye-tracking setups, making HL2 a portable and easily deployable option for hospitals, seamlessly integrating into existing healthcare systems.

\textbf{Contribution:} This research focuses on developing a Mixed Reality-based system for tracking and evaluating eye movements. We present a medical scenario and detail the creation of an application designed to capture eye-tracking signals through Mixed Reality technology for neurodegenerative disease assessment. Additionally, we introduce a pipeline for extracting clinically relevant features from eye-gaze analysis, emphasizing the system's medical capabilities. The study included both healthy individuals and patients with Parkinson's disease, illustrating the viability and promise of our technology in non-intrusively monitoring eye movement patterns. This suggests potential applications for the diagnosis of neurodegenerative diseases.



\section{Methods}

\begin{figure}
    \centering
    \includegraphics[width = 0.45\textwidth]{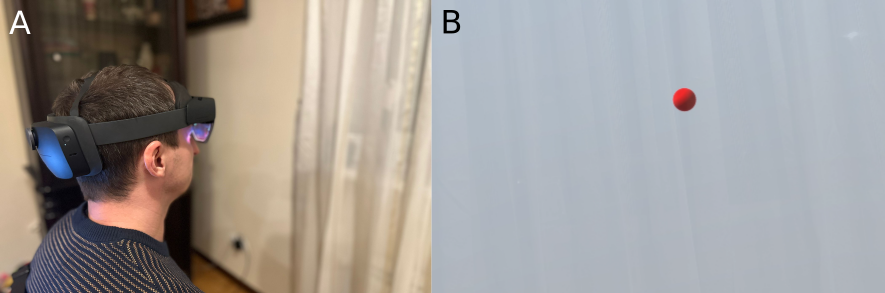}
    \caption{A user engaged in an eye movement task (A), user view during the activity (B).}
    \label{saccades}
\end{figure}

The study involves participants performing a series of four tasks, utilizing an integration of MR glasses to conduct a detailed examination of rapid and deliberate eye movements. Before each task, the user is presented with either textual and auditory instructions or undergoes a brief training session guiding them on the correct task execution. In Figure \ref{saccades}, the user is depicted performing an eye movement task alongside their corresponding perspective. 

MR technology enables to capture and analyze fast eye movements, particularly reflex saccades triggered by external stimuli. The advanced tracking capabilities of the MR glasses facilitate a nuanced exploration of voluntary gaze redirection in antisaccades, emulating complex decision-making scenarios with high fidelity. Moreover, the study delves into memory-guided saccades, wherein participants navigate an MR environment, providing insights into the interplay between memory processes and ocular motor control. The application of MR glasses extends to examining slow eye movements, including gaze fixation and smooth pursuit. Gaze fixation is scrutinized in dynamic real-world scenarios, allowing for a fine-grained visual attention analysis during the task.

\subsection{Acquisition Procedure}
In the first task, participants focus on appearing points, each preceded by a central point, and perform 30 repetitions. Points randomly appear at positions (-20, -10, 10, and 20 degrees relative to the participant's head) and are displayed at continuous random intervals ranging from 1 to 3 seconds.

The second task, known as anti-saccades, is structurally identical (handled by the same code segment), with the difference being in the instruction. Participants are asked to direct their gaze in the opposite direction to the appearing point, so if the point appears on the right side, the participant should look to the left.

The third task called memory-guided saccades, is presented as a challenging task, preceded by instructions and an exercise with a guiding arrow that is later absent in the main part of the task. The instructional phase involves 10 repetitions, and participants are instructed to look at the location where the point appeared previously. The sequence and placement of points are analogous to the previous tasks, and the times and places of their appearance are also randomized. In the main part of the task, 30 repetitions are performed.

In the fourth task, involving eye-tracking movements, participants are tasked with observing a moving point. Points oscillate from -15 to 15 degrees and back relative to the participant's head. Thirty repetitions are performed, and two frequencies are randomly selected: 0.2 or 0.4 Hz.
   
\subsection{Technical Details}

The HoloLens 2 features an integrated eye tracker with a sampling rate of 30 Hz and a spatial accuracy of 1.5° \cite{HL_source}, capable of concurrently recording gaze data from both eyes. Although the standard eye-tracking API of the device does not permit the extraction of raw monocular signals, such functionality is accessible through the Eye Tracking Extended API. Our contribution lies in the development of an application for the acquisition of eye-tracking data, utilizing the standard 30 Hz eye-tracking capabilities of the HoloLens Glasses. This frequency is commonly employed in commercial Mixed Reality systems not intended for scientific research and is readily accessible to software developers. We aimed to examine the feasibility of utilizing eye tracking technology, even at lower frequencies, which is standardly available in MR solutions, for the detection of syndromes associated with neurodegenerative diseases. The application is composed of two layers: an interactive app built on Unity for the presentation of tasks and a C++ based layer for the low-level acquisition of sensor data. This setup enables the collection of signals from all available sensors, including the inertial measurement unit (IMU), microphones, and eye-tracking data (such as the direction of eye gaze and gaze distance), along with acquisition timestamps. The Unity layer similarly records the positions of holographic objects presented to the user. All collected data is stored in the internal memory of the HoloLens 2 and processed post-procedure.

\subsection{Data Processing} \label{data_processing_section}
Raw data collected from HL2 including an eye tracker timestamp, the eye gaze vector, and the eye gaze origin point, were extracted. The eye gaze position was calculated as $origin + direction$ vector sum. The X-axis component of the eye gaze position was extracted and normalized to the HL2 field of view. This normalization facilitates comparison with the positions of visual stimuli. The positions of these visual stimuli were obtained from the logging component of the HL2 Unity application. The stimulus display timestamp and the X-axis position component were extracted from this data. The eye gaze position data underwent filtering using a Savitzky-Golay filter \cite{savgol} for noise reduction and smoothing. In the sections focusing on reflex, antisaccades, and memory saccades, saccade points in the filtered signal were identified as the first gaze data point with an amplitude exceeding half of the signal's standard deviation, in response to a peripheral visual stimulus. If no such response occurred before the appearance of the next visual stimulus, no saccade point was recorded.

For reflex saccades 4 parameters were extracted:
\paragraph{Latency}
Latency was calculated as a difference between the timestamps of a saccade point, and the first visual stimulus displayed before that saccade point. Due to how saccade points were registered, this guaranteed no instances of mismatching stimulus-saccade pairs were included.
\paragraph{Saccade speed}
Due to a limiting sampling rate of 60 Hz, the saccade speed could only be approximated. The gradient of the gaze signal was calculated, and filtered once again with a Savitzky-Golay filter \cite{savgol}. For each registered saccade, the maximum value in a radius of 20 samples from the saccade point was recognized as a speed approximation for that saccade. 
\paragraph{Average amplitudes}
For 10\textdegree\ and 20\textdegree\ peripheral visual stimuli, an actual gaze amplitude was approximated. First, stimuli signals were resampled to gaze signal sampling times. Then, parts of the gaze signal corresponding to 10\textdegree\ and 20\textdegree\ visual stimuli were extracted and averaged. 
\paragraph{Average fixation time}
For each registered saccade, fixation time was calculated. As with saccade speed, the gradient of the gaze signal was calculated and filtered. For threshold calculation, outliers were removed by a modified Z-score as described by Iglewicz \cite{iglewicz_outliers}. The gaze stabilization threshold was then calculated as a 0.05 of the standard deviation of the outlier-removed gradient signal. \newline
For each saccade point, the initial start of a fixation was identified as a timestamp of the first gradient datapoint below that threshold (and after the saccade); and the end of a fixation as the next gradient datapoint exceeding the threshold. To eliminate microfixations, any fixations less than 50 ms in length were discarded, and the above calculation was repeated, starting from the end of discarded microfixation (instead of a saccade point). If no fixation was detected before the next visual stimulus, the longest of microfixations was registered for calculation.  
The parameters used to characterize the antisaccades included:
\paragraph{Average latency}
Average latency was calculated as a difference between a timestamp of first saccade in a correct direction in response to a stimulus, and corresponding stimulus.
\paragraph{Incorrect saccades ratio}
Instances of stimuli where the first registered saccade was in the same direction as the stimuli were counted as incorrect and a ratio of incorrect to incorrect summed with correct ones was calculated.

The parameters for memory-guided saccades are as follows:
\paragraph{Incorrect saccades ratio}
Instances of stimuli where the first registered saccade was in a different direction to a previous stimulus were counted as incorrect, and a ratio of incorrect to incorrect summed with correct results was calculated. 

In computing smooth pursuit, the algorithm considers:
\paragraph{Smooth pursuit speed}
The speed of the gaze signal was calculated as a gradient of the absolute value of gaze data. Mean and standard deviation were evaluated.
\paragraph{Smooth pursuit acceleration}
Acceleration of the gaze signal was calculated as a second derivative of gaze data. RMS of acceleration was evaluated.

Parameters calculated from each part of the experiment are as follows: 
\begin{itemize}
    \item Reflex Saccades: average latency [ms], average saccade speed [\textdegree/$ms$], average amplitudes for 10 and 20 degrees [\textdegree], average fixation time [ms], 
    \item Anti Saccades: average latency [ms], incorrect saccades ratio, 
    \item Memory Saccades: incorrect saccade ratio, 
    \item Smooth Pursuit: average saccade speed [\textdegree/$ms$], root mean square of acceleration [\textdegree/$ms^2$], speed standard deviation.
\end{itemize}

        



\section{Results}
\subsection{Preprocessed data}
\begin{figure}[tbh!]
    \centering
    \includegraphics[width = 0.5\textwidth]{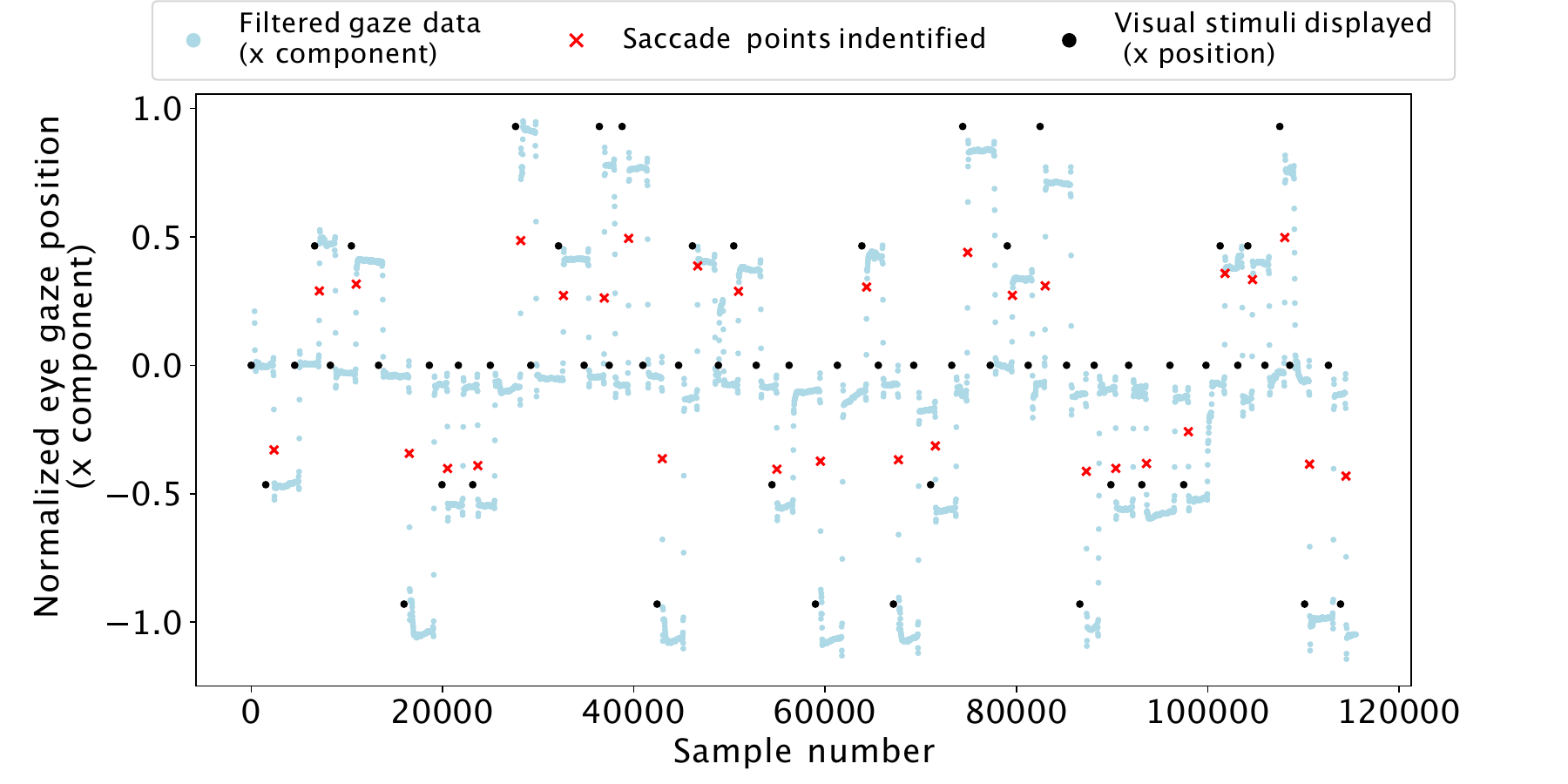}
    \caption{Example of preprocessed signal from reflex saccades part.}
    \label{example_preprocessed}
\end{figure}

Raw data from Hololens 2 device was received as a binary file. The file was parsed to an array of structures containing both head and eye movement data. From that structure, fields representing timestamps, eye gaze vectors, and eye gaze origin point were extracted. The horizontal axis component was extracted, normalized and filtered as described in section \ref{data_processing_section}. Study information data (information about visual stimulus displayed with corresponding timestamp) was transformed to an array of tuples (timestamp, x-axis position). Position along the x-axis was normalized in the same way as eye gaze data, thus enabling comparisons. An example of the results of initial data preprocessing is presented in figure \ref{example_preprocessed}.
\subsection{Case study}
\begin{figure}
    \centering
    \includegraphics[width = 0.48\textwidth]{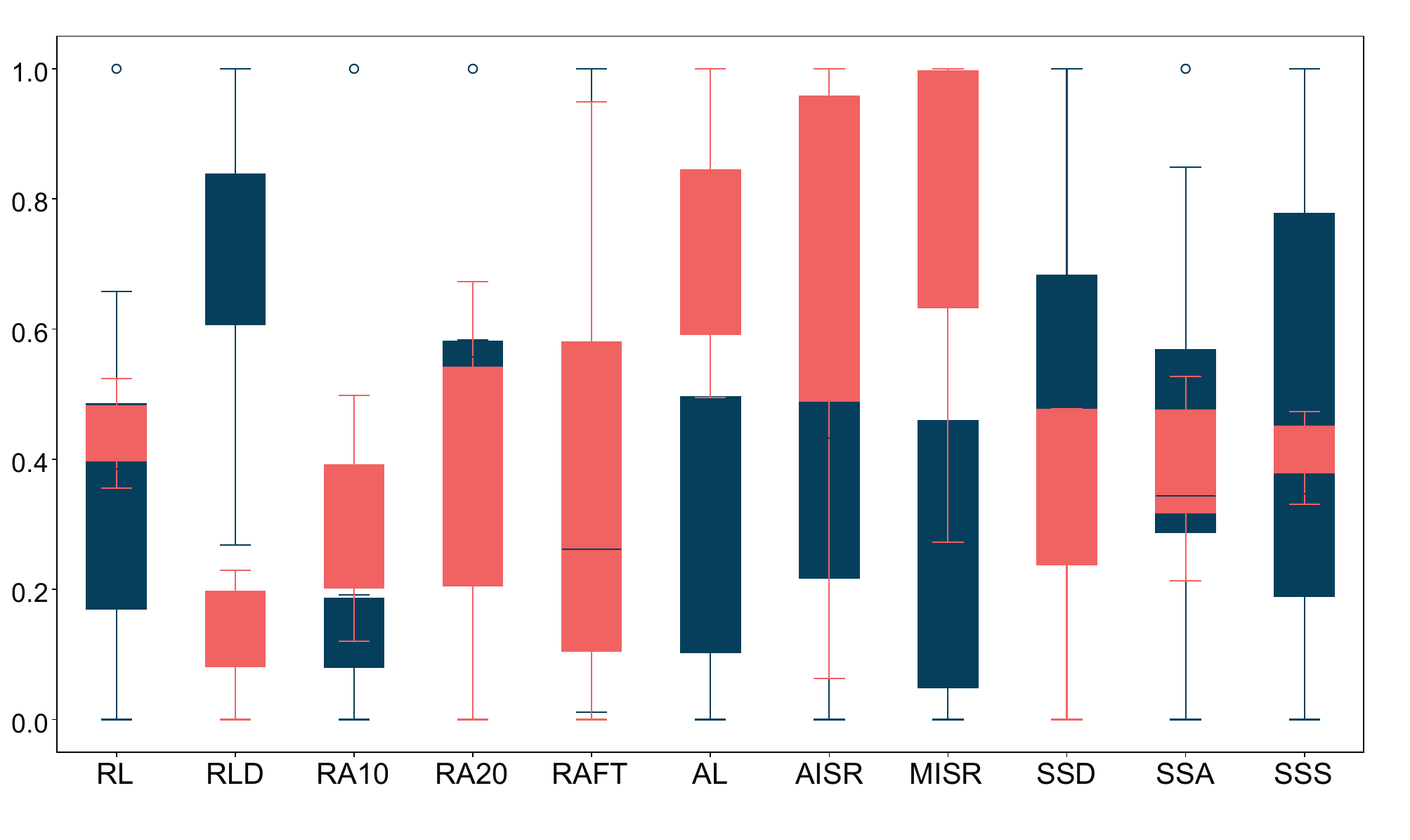}
    \caption{Boxplots for parameters for healthy group (dark blue) and PD group (light red). Depicted parameters: \textbf{RL} - reflex latency, \textbf{RSD} - reflex speed, \textbf{RA10} - reflex amplitude (10\textdegree), \textbf{RA20} - reflex amplitude (20\textdegree), \textbf{RAFT} - reflex average fixation time, \textbf{AL} - anti-latency, \textbf{AISR} - anti-incorrect saccades ratio, \textbf{MISR} - memory-incorrect saccades ratio, \textbf{SSD} - smooth pursuit speed, \textbf{SSA} - smooth pursuit acceleration RMS, \textbf{SSS} - smooth pursuit std. deviation}
    \label{boxplots}
\end{figure}

For a section of study participants, parameters were normalized to [0,1] and averaged over the healthy (n=13) and neurodegenerative (PD) (n=4) groups. Boxplots of those parameters are displayed in figure \ref{boxplots}. Clear disparities between healthy controls and Parkinson's disease (PD) patients are evident in reflex speed and reflex amplitude at 20 degrees. In contrast, the interquartile ranges vary between the analyzed groups for the remaining parameters, while the max-min ranges may lead to overfitting.
.



\section{Discussion and Conclusions}

The study demonstrates the potential of using the commercially available MR glasses for identifying symptoms associated with neurological disorders, particularly Parkinson's Disease. By leveraging advanced eye-tracking technology, the device enables the detection of characteristic eye movement patterns, such as reflexive saccades, antisaccades, memory-guided saccades, and smooth pursuit. The findings align with existing literature on eye movement abnormalities in PD. The study specifically highlights the hypometric nature of reflexive saccades in PD, difficulties in antisaccade tasks, impairment in memory-guided saccades, and reduced smooth pursuit, providing validation for the use of HL2 in monitoring these symptoms. The study emphasizes the generalizability of the approach, considering the 30 Hz eye-tracking frequency commonly found in non-scientific, low-cost, and gaming appliances. The ability to detect neurological symptoms with such simple and accessible technology holds promise for lowering the cost of diagnosis and increasing the adoption of eye-tracking technology in healthcare. The study recognizes the prevalence of the 30 Hz eye-tracking frequency in non-scientific, low-cost, and gaming appliances. The ability to detect meaningful insights into neurological conditions using such accessible technology suggests a promising avenue for lowering diagnosis costs and enhancing the widespread adoption of eye-tracking technology in healthcare. This finding underscores the potential impact of integrating affordable devices with simpler specifications into routine clinical assessments, particularly in resource-constrained environments where cost-effective solutions are crucial \cite{eyetracking_meta_quest_1}. This has the potential to benefit a larger population, particularly in resource-constrained environments where access to high-end diagnostic equipment may be limited \cite{arpaia_design_2021}. The study's contribution lies in its assessment of the HL2 glasses for eye-tracking, specifically in detecting eye movements indicative of neurodegenerative diseases. Future directions may involve expanding the study to a larger participant pool, incorporating longitudinal data to observe the progression of symptoms, and refining the methodology based on feedback and advancements in technology.

\vspace{12pt}

\bibliographystyle{IEEEbib}
\bibliography{bibliography}

\begin{thebibliography}{10}

\bibitem{pd10}
{Parkinson's Foundation},
\newblock ``{Parkinson Statistics},'' {https://www.parkinson.org/understanding-parkinsons/statistics }, 2018,
\newblock Online; accessed 26 January 2024.

\bibitem{sun_monitoring_2023}
Yue~Ran Sun, Sinem~B Beylergil, Palak Gupta, Fatema~F Ghasia, and Aasef~G Shaikh,
\newblock ``Monitoring eye movement in patients with parkinson’s disease: What can it tell us?,''
\newblock vol. Volume 15, pp. 101--112.

\bibitem{jung_abnormal_2019}
Ileok Jung and Ji-Soo Kim,
\newblock ``Abnormal eye movements in parkinsonism and movement disorders,''
\newblock vol. 12, no. 1, pp. 1--13.

\bibitem{alfalahi_scoping_2023}
Hessa Alfalahi, Sofia~B. Dias, Ahsan~H. Khandoker, Kallol~Ray Chaudhuri, and Leontios~J. Hadjileontiadis,
\newblock ``A scoping review of neurodegenerative manifestations in explainable digital phenotyping,''
\newblock vol. 9, no. 1, pp. 1--22,
\newblock Number: 1 Publisher: Nature Publishing Group.

\bibitem{krichen_anomalies_2021}
Moez Krichen,
\newblock ``Anomalies detection through smartphone sensors: A review,''
\newblock vol. 21, no. 6, pp. 7207--7217.

\bibitem{aarsland_parkinson_2021}
Dag Aarsland, Lucia Batzu, Glenda~M. Halliday, Gert~J. Geurtsen, Clive Ballard, K.~Ray~Chaudhuri, and Daniel Weintraub,
\newblock ``Parkinson disease-associated cognitive impairment,''
\newblock vol. 7, no. 1, pp. 1--21,
\newblock Number: 1 Publisher: Nature Publishing Group.

\bibitem{motahari-nezhad_digital_2022}
Hossein Motahari-Nezhad, Meriem Fgaier, Mohamed~Mahdi Abid, Márta Péntek, László Gulácsi, and Zsombor Zrubka,
\newblock ``Digital biomarker–based studies: Scoping review of systematic reviews,''
\newblock vol. 10, no. 10, pp. e35722,
\newblock Company: {JMIR} {mHealth} and {uHealth} Distributor: {JMIR} {mHealth} and {uHealth} Institution: {JMIR} {mHealth} and {uHealth} Label: {JMIR} {mHealth} and {uHealth} Publisher: {JMIR} Publications Inc., Toronto, Canada.

\bibitem{gsaxner_hololens_2023}
Christina Gsaxner, Jianning Li, Antonio Pepe, Yuan Jin, Jens Kleesiek, Dieter Schmalstieg, and Jan Egger,
\newblock ``The {HoloLens} in medicine: A systematic review and taxonomy,''
\newblock vol. 85, pp. 102757.

\bibitem{orlosky_emulation_2017}
Jason Orlosky, Yuta Itoh, Maud Ranchet, Kiyoshi Kiyokawa, John Morgan, and Hannes Devos,
\newblock ``Emulation of physician tasks in eye-tracked virtual reality for remote diagnosis of neurodegenerative disease,''
\newblock vol. 23, no. 4, pp. 1302--1311.

\bibitem{kourtis_digital_2019}
Lampros~C. Kourtis, Oliver~B. Regele, Justin~M. Wright, and Graham~B. Jones,
\newblock ``Digital biomarkers for alzheimer’s disease: the mobile/wearable devices opportunity,''
\newblock vol. 2, no. 1, pp. 9.

\bibitem{HL_source}
{Eye tracking on HoloLens 2},
\newblock ``{Microsoft},'' {https://learn.microsoft.com/en-us/windows/mixed-reality/design/eye-tracking }, 2023,
\newblock Online; accessed 26 January 2024.

\bibitem{savgol}
Abraham. Savitzky and M.~J.~E. Golay,
\newblock ``Smoothing and differentiation of data by simplified least squares procedures.,''
\newblock {\em Analytical Chemistry}, vol. 36, no. 8, pp. 1627--1639, 1964.

\bibitem{iglewicz_outliers}
Boris Iglewicz and David Hoaglin,
\newblock {\em How to Detect and Handle Outliers},
\newblock American Society for Quality, 1993.

\bibitem{eyetracking_meta_quest_1}
Shogo Shimada, Yasushi Ikei, Nobuyuki Nishiuchi, and Vibol Yem,
\newblock ``Study of cybersickness prediction in real time using eye tracking data,''
\newblock in {\em 2023 IEEE Conference on Virtual Reality and 3D User Interfaces Abstracts and Workshops (VRW)}, 2023, pp. 871--872.

\bibitem{arpaia_design_2021}
Pasquale Arpaia, Egidio De~Benedetto, and Luigi Duraccio,
\newblock ``Design, implementation, and metrological characterization of a wearable, integrated {AR}-{BCI} hands-free system for health 4.0 monitoring,''
\newblock vol. 177, pp. 109280.

\end{thebibliography}

\end{document}